\documentclass[aps,prl,twocolumn,amsmath,amssymb,nofootinbib,superscriptaddress]{revtex4-2}

\usepackage{graphicx}
\usepackage{xcolor}
\usepackage[colorlinks=true,linkcolor=blue!55!black,citecolor=blue!55!black,urlcolor=blue!55!black]{hyperref}

\newcommand{\kB}{k_{B}}
\newcommand{\ttr}{\tau_{\rm tr}}
\newcommand{\aF}{\alpha^{2}F}
\renewcommand{\Im}{\operatorname{Im}}
\renewcommand{\Re}{\operatorname{Re}}
\newcommand{\Asq}{A_{1}^{\square}}

\begin{document}

\title{Planckian dissipation per unit critical entropy as a coupling-free probe of the strange-metal bath}

\author{Dino \v{S}krgi\'c}
\email{dinoskrg@buffalo.edu}
\affiliation{Department of Physics, University at Buffalo, SUNY, Buffalo, NY 14260, USA}

\author{Dejan Stojkovic}
\email{ds77@buffalo.edu}
\affiliation{Department of Physics, University at Buffalo, SUNY, Buffalo, NY 14260, USA}

\date{\today}

\begin{abstract}
Strange metals pair a resistivity linear in temperature with a specific heat $C/T$ rising logarithmically on cooling. We find that, if electrons scatter off a local critical bath, the coupling cancels in the ratio $Q$ of their coefficients, leaving two measurable numbers: how slowly the bath relaxes, seen in photoemission or optics, and the share of critical entropy that relaxes current. Baths form classes: $Q\le4.91$ for a conformal bath, as SYK models confirm; 9--15 for the Yukawa--SYK field; and far less for Kondo lattices. LSCO, at $Q\approx9$, exceeds the conformal bound, pointing to the Yukawa--SYK field.
\end{abstract}

\maketitle


\textit{Introduction.}---Near a quantum critical point, strange metals combine a resistivity $\rho=\rho_{0}+A_{1}T$, a transport rate close to the Planckian value $\hbar/\ttr\sim\kB T$~\cite{Bruin2013,Legros2019,HartnollMackenzie2022}, and an electronic specific heat $C_{\rm el}/T=B\ln(T_{0}/T)$. In hole-doped cuprates at the end of the pseudogap (doping $p^{*}$), these laws hold down to the lowest temperatures measured once superconductivity is suppressed by a strong magnetic field~\cite{Daou2009,Cooper2009,Legros2019,Michon2019,Girod2021}, and the $T$-linear rate is the same all around the Fermi surface, with no hot spots~\cite{Grissonnanche2021}. Heavy-fermion critical points show the same pair of laws~\cite{Lohneysen2007,Custers2003,Bianchi2003,TaupinPaschen2022}.

\begin{figure}[t]
\includegraphics[width=\columnwidth]{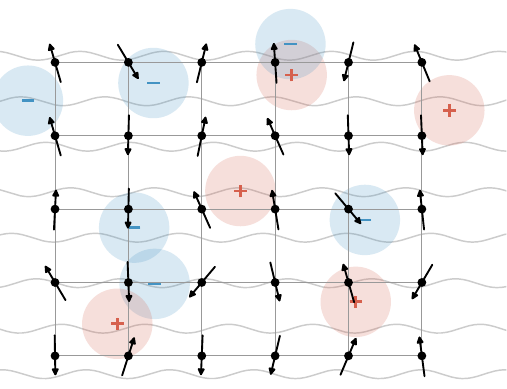}
\caption{A schematic of the strange-metal bath matching the cuprate data. Local units (arrows) on a two-dimensional lattice, with nearly fixed amplitude and critical at $p^{*}$, sit in the sea of conduction electrons (grey lines). They couple to the electron density only through spatially random local fields of zero mean (red $+$, blue $-$), so each electron scatters locally and the transport rate is isotropic. The electrons in turn damp the units, $D^{-1}=m+\nu_{0}q^{2}+\gamma|\Omega|$, and the critical tuning leaves a thermal mass $m=r\gamma\kB T$ with $r\approx0.3$--$0.5$, so the bath relaxes two to three times more slowly than a conformal one ($r\approx1.1$). This is the two-dimensional Yukawa--SYK field~\cite{Esterlis2021,PatelGuoEsterlisSachdev2023}, the slow bath that the data favor. The measurements outlined in Fig.~\ref{fig:map} would confirm or exclude it.}
\label{fig:lattice}
\end{figure}

The marginal Fermi liquid~\cite{Varma1989} and its modern descendants reproduce both laws~\cite{SachdevYe1993,Kitaev2015,SongJianBalents2017,Chowdhury2018,ChowdhuryRMP2022,FaulknerLiuMcGreevyVegh2011,Esterlis2021,PatelGuoEsterlisSachdev2023,SachdevReview2025} but they disagree on what the electrons scatter from: islands of strongly interacting fermions, a holographic critical sector, or a critical boson with spatially random couplings. Each theory has a free coupling, so the Planckian coefficient alone cannot classify them. The specific heat and the transport rate have been compared at a common coupling~\cite{Varma2020,Esterlis2021,PatelGuoEsterlisSachdev2023,Michon2023}, and the random-coupling theory reproduces key features of the optical conductivity of La$_{2-x}$Sr$_{x}$CuO$_{4}$ (LSCO)~\cite{Li2024}.

\begin{figure}[t]
\includegraphics[width=\columnwidth]{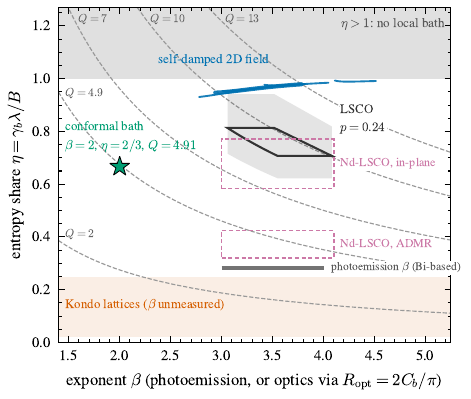}
\caption{Classifying strange-metal baths without the coupling. Horizontal: how slowly the bath relaxes, $\beta$ [Eq.~(\ref{eq:beta})], read from nodal photoemission or from the optical ratio $R_{\rm opt}=2C_{b}/\pi$. Vertical: the share $\eta=\gamma_{b}\lambda/B$ of the critical entropy carried by current-relaxing scattering, mapping distinct baths onto a universal diagram. Dashed: contours of the coupling-free ratio of resistivity slope to specific-heat logarithm, $Q=\alpha\gamma_{b}/B=C_{b}(\beta)\,\eta$, with $C_{b}(\beta)$ of $w=(2/\pi)\arctan(x/r)$ (other shapes differ by at most $5\%$).
Grey above $\eta=1$: not reached by a fixed bath with $w\le1$ or by the self-damped field. Star: the conformal bath, reproduced to $0.2\%$ by two-band SYK. Blue lines: the self-damped field, $\eta=1/(1+\delta)$ along the large-$M$ saddle for $\gamma\kB T/E_{L}=10^{-5}$--$10^{-2}$, $u\rho=0.3$, $1$ and $\infty$ and $E_{c}=0.3$--$3E_{L}$ (thick: $u\rho=\infty$, $E_{c}=E_{L}$). Grey region: LSCO at $p=0.24$ from published optics, resistivity and specific heat with the band of Ref.~\cite{Michon2023}, for $\alpha=5$--$5.7$, $\lambda=0.40$--$0.46$ and $B=2.2\pm0.3$~mJ\,K$^{-2}$\,mol$^{-1}$ (outline: $B=2.2$); since $C_{b}=\alpha/\lambda$ and $\eta\propto\lambda$, it runs along contours of $Q$. Bar: $\beta=3$--$4$ from photoemission in Bi-based cuprates (height arbitrary). Dashed boxes: Nd-LSCO at $p=0.24$ from the in-plane and the ADMR rates, with LSCO's $\beta$ assumed. Orange: four Kondo lattices, $\beta$ unmeasured.}
\label{fig:map}
\end{figure}

In this Letter we remove the free coupling for a local and marginal bath, which leaves us with an exponent $\beta$ that says how slowly the bath relaxes, seen by photoemission and optics, and the share of the critical entropy carried by processes that relax the current, $\eta$. We find their product to be the coupling-free ratio $Q$ of the resistivity slope to the specific-heat logarithm, up to a known function of $\beta$. We then compute the numbers for two relevant models of strange metal baths, a bath conformal in time and a self-damped two-dimensional field. We also place published cuprate data on a $(\beta,\eta)$ diagram (Fig.~\ref{fig:map}), and identify the measurements that would decide.

\textit{A local marginal bath.}---A boson scattering electrons by small momenta gives the right self-energy (marginal) but not a $T$-linear resistivity in two dimensions. On the other hand, a boson with finite wave vector creates hot spots, which angle-dependent magnetoresistance (ADMR) does not see~\cite{Hertz1976,Millis1993,HlubinaRice1995,DellAnnaMetzner2007,MaslovYudsonChubukov2011,Grissonnanche2021}. A bath that couples to the electrons locally (Fig.~\ref{fig:lattice}) avoids both, whether or not the correlation length diverges. SYK island lattices~\cite{SongJianBalents2017,Chowdhury2018,PatelMcGreevyArovasSachdev2018}, semi-local holographic metals~\cite{FaulknerLiuMcGreevyVegh2011,IqbalLiuMezei2012} and critical bosons with random-in-space couplings~\cite{Esterlis2021,PatelGuoEsterlisSachdev2023,Li2024} are some realizations. For a momentum-independent vertex and self-energy, current-vertex corrections vanish and the transport rate is a thermal average of the single-particle rate~\cite{PatelMcGreevyArovasSachdev2018}. We therefore take a local, marginal Eliashberg function, exact at large-$N$ for random couplings~\cite{SachdevYe1993,Kitaev2015,MaldacenaStanford2016,EsterlisSchmalian2019,Esterlis2021,PatelGuoEsterlisSachdev2023,ChowdhuryRMP2022},
\begin{equation}
\aF(\Omega,T)=\frac{\lambda}{2}\,w\!\left(\frac{\Omega}{\kB T}\right),\qquad w(x\gg 1)\to1,
\label{eq:w}
\end{equation}
which depends on $\Omega/\kB T$ alone because criticality makes $T$ the only scale. It is flat above $\kB T$, and its shape $w$ says how the bath switches on below it: a bath that relaxes at a rate $\Gamma=r\kB T$ has $w$ reaching half strength near $\Omega\approx r\kB T$.

We write the transport rate per band mass as $\hbar/\ttr=\alpha\kB T$; the Planckian coefficient of Ref.~\cite{Legros2019} is its mass-renormalized counterpart, $\alpha^{*}=\alpha\,m_{b}/m^{*}$. Photoemission measures the shape directly (Sec.~\ref{sec:Sbath}): $-\Im\Sigma\to\frac{\pi}{2}\lambda|\varepsilon|$ for $|\varepsilon|\gg\kB T$, while at the Fermi level $-\Im\Sigma(0,T)=\frac{\pi}{2}\lambda\beta\kB T$ with
\begin{equation}
\beta=2\int_{0}^{\infty}\frac{w(x)}{\sinh x}\,dx .
\label{eq:beta}
\end{equation}
A slower bath has more thermally occupied modes below $\kB T$ and therefore a larger $\beta$. This is the exponent fitted in nodal photoemission, $\Im\Sigma\propto[\varepsilon^{2}+(\beta\kB T)^{2}]^{1/2}$~\cite{Reber2019,Smit2021}.

Transport sees the same shape through $\alpha=C_{b}[w]\,\lambda$, where $1/C_{b}$ is the Kubo average of $\lambda\kB T/2|\Im\Sigma|$. For every family of shapes we examined, $C_{b}/\beta=3.48$--$3.76$ when $\beta\ge1.5$, so $C_{b}\simeq3.6\beta$ to $5\%$ (Sec.~\ref{sec:Sbath}). Optics gives $C_{b}$ without $\lambda$ or the spectral weight, because the zero-frequency optical rate over its high-frequency slope is $R_{\rm opt}=2C_{b}/\pi\simeq2.3\beta$.

The entropy determines where the logarithm sits. The electrons' mass enhancement, $1+\lambda\ln(\omega_{c}/|\varepsilon|)$, contributes $\gamma_{b}\lambda$ to $B$, where $\gamma_{b}$ is the band Sommerfeld coefficient, and entropy carried by the bath itself adds to it. and entropy carried by the bath itself adds to it. We define $b\equiv B/(\gamma_{b}\lambda)$ and $\eta\equiv1/b$, the share of the critical entropy carried by the processes that relax current. A fixed bath whose spectrum rises to its plateau from below has $b\ge1$, and so does the self-damped field discussed below. The coupling then cancels in the ratio
\begin{equation}
Q\equiv\frac{\alpha\,\gamma_{b}}{B},
\label{eq:Qdef}
\end{equation}
which becomes
\begin{equation}
Q=C_{b}[w]\,\eta\simeq3.6\,\beta\,\eta,\qquad\eta\le1.
\label{eq:Q}
\end{equation}
Equation~(\ref{eq:Q}) is an identity once $\eta$ is defined; its content is that $C_{b}$ is fixed by a measurable exponent and that $\eta$ takes definite values for physically distinct baths. Given $\lambda$, only two of $Q$, $\beta$ and $\eta$ are independent, so Fig.~\ref{fig:map} uses $\beta$ and $\eta$ as axes and shows $Q$ as a family of contours, the combination that needs neither $\lambda$ nor spectroscopy. Locality itself is tested between probes: a local self-energy requires $R_{\rm opt}\simeq2.3\beta$.

\textit{Classifications.}---We first turn to conformal baths. SYK islands, AdS$_{2}$ holography and other baths with emergent time-reparametrization symmetry have $w=\tanh(x/2)$~\cite{SachdevQPT,ParcolletGeorges1999,PatelMcGreevyArovasSachdev2018}, so $\beta=2$ and $C_{b}=2\pi^{3}/7\zeta(3)=7.370$, found numerically in Ref.~\cite{PatelMcGreevyArovasSachdev2018} and in closed form in Ref.~\cite{MaebashiVarma2025}. The bath's own temperature-dependent shape carries entropy: removing spectral weight below $\kB T$ adds $3I_{w}/\pi^{2}$ to $b$, with $I_{w}=\int_{0}^{\infty}x(1-w)\,dx=\pi^{2}/6$ (Sec.~\ref{sec:Sbath}). Hence $b=\frac32$, $\eta=\frac23$, and
\begin{equation}
Q_{\rm conf}=\frac{4\pi^{3}}{21\zeta(3)}=4.913 .
\label{eq:Qconf}
\end{equation}
For a flat density of states the free energy is exactly linear in $\lambda$, so Eq.~(\ref{eq:Qconf}) holds to all orders; the large-$N$ two-band SYK model~\cite{Chowdhury2018} gives $Q/Q_{\rm conf}=1.0020$--$1.0024$ over a ninefold range of coupling (Sec.~\ref{sec:Sbath}). The value $\eta=\frac23$ corrects the usual identification of the logarithm in $C/T$ with the mass logarithm~\cite{CrisanMoca1996,PatelMcGreevyArovasSachdev2018}, because a third of it is in the bath. Entropy carried by anything that does not relax current only adds to $B$, so a conformal bath has $Q\le Q_{\rm conf}$, an upper bound testable with transport and calorimetry alone.

Consider now the opposite case, a field damped by the same electrons it scatters. We take an $M$-component two-dimensional overdamped field, $D^{-1}=m(T)+\nu_{0}q^{2}+\gamma|\Omega|$, coupled to the local density through random vertices of zero mean and variance $\overline{g'^{2}}$, as in the two-dimensional Yukawa--SYK theory~\cite{Esterlis2021,PatelGuoEsterlisSachdev2023,SachdevReview2025}. Its local spectrum is $w=(2/\pi)\arctan(x/r)$ with $r=m/\gamma\kB T$. One vertex sets both the electrons' coupling, $\lambda=M\overline{g'^{2}}N_{0}\rho$, and the field's Landau damping, $\gamma=2\pi\overline{g'^{2}}N_{0}^{2}$, with $N_{0}$ and $\rho$ the electron and field densities of states. Because $m$ is stationary at the large-$M$ saddle, $S=-\partial F/\partial T$ at fixed $m$, and for $\gamma\kB T$ well below the field's bandwidth $E_{L}$ the field's determinant gives (Sec.~\ref{sec:Sfield}; $\hbar=\kB=1$)
\begin{equation}
S_{\phi}=\frac{M\rho\gamma T}{\pi}\left[\frac{\pi^{2}}{3}\ln\frac{E_{L}}{\gamma T}-K(r)\right],
\label{eq:Sfield}
\end{equation}
with $K(r)$ regular. $S_{\phi}$ vanishes as $T\to0$, leaving no residual entropy, and its leading logarithmic coefficient, $B=\pi M\rho\gamma/3$, equals $\gamma_{b}\lambda$ exactly. Hence $b=1$: the logarithm in $C/T$ is the electrons' mass logarithm, consistent with the whole enhancement of $C/T$ found in the boson determinant in Ref.~\cite{Esterlis2021}. The temperature dependence of $r$ adds $\delta=b-1=0.007$--$0.074$, so $\eta=0.93$--$0.99$. A fixed-bath formula applied to the arctan spectrum would instead give a temperature-independent entropy that diverges with the cutoff; that piece is the term linear in $m$, which the critical tuning of the bare mass cancels.

\begin{figure}[t]
\includegraphics[width=\columnwidth]{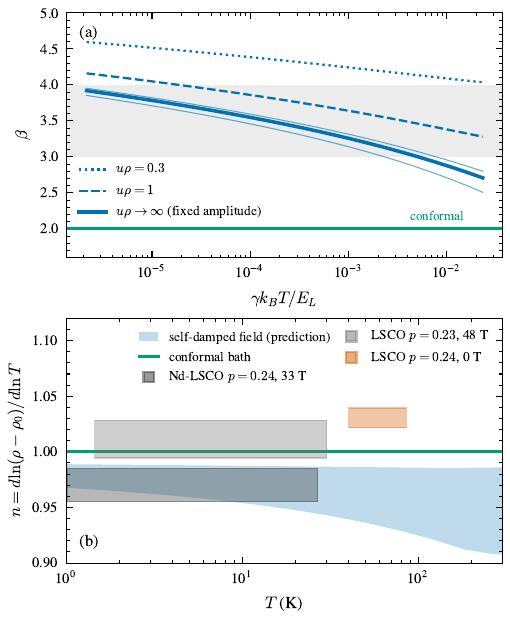}
\caption{The self-damped field. (a)~$\beta$ from the large-$M$ saddle against $\gamma\kB T/E_{L}$ for $u\rho=0.3$ (dotted), $1$ (dashed) and $\infty$ (fixed amplitude, thick) at frequency cutoff $E_{c}=E_{L}$; thin: fixed amplitude at $E_{c}=0.3E_{L}$ and $3E_{L}$. The fixed-amplitude limit is a floor. Grey band: nodal photoemission in Bi$_{2}$Sr$_{2}$CaCu$_{2}$O$_{8+\delta}$ and Bi$_{2}$Sr$_{2}$CuO$_{6+\delta}$~\cite{Reber2019,Smit2021}. Green: the conformal bath. (b)~The drift test. Blue band: the local exponent $n=d\ln(\rho-\rho_{0})/d\ln T$ predicted from $\rho-\rho_{0}\propto T\,C_{b}(T)$, for the parameters of (a) and $E_{L}/\gamma=1$--$10$~eV. Green: the conformal bath, $n=1$. Boxes: power-law exponents fitted to published resistivities over the window shown, with their spread across windows or $\pm1$ standard deviation: Nd-LSCO, $p=0.24$, $33$~T~\cite{Daou2009,Legros2019}; LSCO, $p=0.23$, $48$~T~\cite{Cooper2009,Legros2019}; and the zero-field LSCO crystal of Fig.~\ref{fig:data}b over the predicted window $40$--$85$~K~\cite{Michon2023}.}
\label{fig:field}
\end{figure}

The field is at its upper critical dimension, $d+z=4$, where $u\phi^{4}$ is marginally irrelevant and the thermal mass is suppressed by a logarithm~\cite{Esterlis2021}. Solving the large-$M$ saddle for $m(T)$ over three decades of both $u$ and $\gamma\kB T/E_{L}$, we find that $r$ grows with $u$ and saturates. At $u\to\infty$ the saddle becomes the constraint $\langle\phi^{2}\rangle_{T}=\langle\phi^{2}\rangle_{0}$ of a fixed-amplitude field, with $r=\pi W_{0}(\ell/\pi)/\ell$ and $\ell=\ln(a_{0}e^{\gamma_{E}}/2\pi\kB T)$~\cite{Esterlis2021}. Within the large-$M$ saddle at fixed $E_{L}/\gamma$ this is the fastest the critical field can relax, so that (Fig.~\ref{fig:field}a)
\begin{equation}
\beta\ge\beta_{\infty}\simeq2.8\text{--}3.8,\qquad \eta=0.93\text{--}0.98,\qquad Q\approx9\text{--}13,
\label{eq:floor}
\end{equation}
for $\gamma\kB T/E_{L}=10^{-5}$--$10^{-2}$, while a finite $u$ raises $\beta$ up to $4.5$, $\eta$ up to $0.99$ and $Q$ up to $15.5$. The slowness is not tuned, since the couplings enter only through a logarithm, but it drifts, with $\beta$ rising on cooling by $0.1$--$0.5$ per decade. With no coupling to adjust, the two limits differ by a factor of at least $1.4$ in both $\beta$ and $\eta$.

\begin{figure}[t]
\includegraphics[width=\columnwidth]{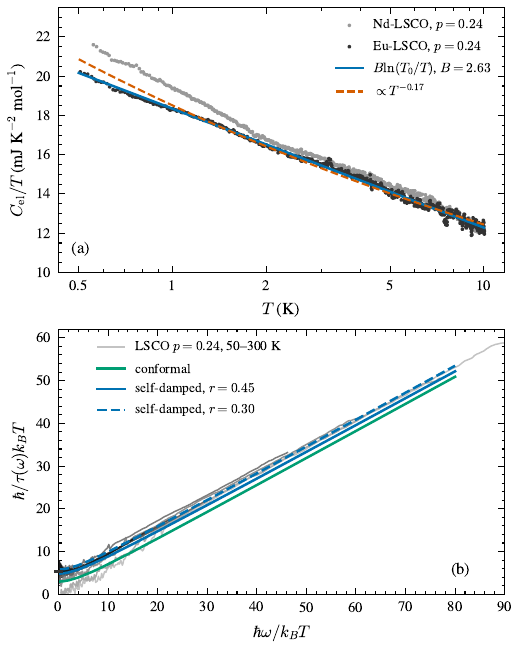}
\caption{Published data at $p=0.24$. (a)~Electronic specific heat of Eu-LSCO (dark) and Nd-LSCO (grey)~\cite{Michon2019}, extracted from the published vector graphics. Blue: logarithmic fit to Eu-LSCO over $0.55$--$10$~K, $B=2.63$~mJ\,K$^{-2}$\,mol$^{-1}$ (r.m.s.\ residual $0.15$); orange dashed: best power law, $\propto T^{-0.17}$ (r.m.s.\ $0.23$). Nd-LSCO has an extra contribution below $2$~K; above $2$~K the two logarithms agree to $6\%$. (b)~Optical scattering rate $\hbar/\tau(\omega)\kB T$ of LSCO at eight temperatures from $50$ to $300$~K~\cite{Michon2023} (grey; vector paths of the published figure, axes calibrated on its frame and ticks), against the local-bath model at the measured slope ($\lambda=0.40$) with the full Kramers--Kronig self-energy: conformal (green) and self-damped field with $r=0.45$ (blue) and $0.30$ (dashed). Only the zero-frequency value distinguishes the models. Black bar: the measured value, $5$ (optics, as quoted in Ref.~\cite{Michon2023}) to $5.7$ (dc).}
\label{fig:data}
\end{figure}

We now place published data on LSCO and its Eu- and Nd-substituted relatives at $p=0.24$ on the map. A local bath with spectrum $\propto\Omega^{s}$ gives $C/T\propto T^{s}$ and $\rho\propto T^{1+s}$ (Sec.~\ref{sec:Sother}). The specific heat of Eu-LSCO~\cite{Michon2019} at $0.55$--$10$~K is fitted better by a logarithm, with $B=2.63$~mJ\,K$^{-2}$\,mol$^{-1}$, than by the best power law, $T^{-0.17}$, which would imply $\rho\propto T^{0.83}$ (Fig.~\ref{fig:data}a). The spectrum is therefore flat, $s=0$, which for an overdamped $z=2$ field requires $d=2$.

The ratio $Q$ needs neither $\lambda$ nor spectroscopy. The optical rate of LSCO at $p=0.24$ shows $\omega/T$ scaling between $50$ and $300$~K, with the zero-frequency value $\alpha=5$ quoted by Michon et al.~\cite{Michon2023}, or $5.7$ from the dc resistivity with their conversion. With their band, $\gamma_{b}=3.89$~mJ\,K$^{-2}$\,mol$^{-1}$, and $B=2.2\pm0.3$~\cite{Girod2021}, $Q=\alpha\gamma_{b}/B=8.8$--$10.1$ ($7.8$--$11.7$ with the error on $B$), about twice $Q_{\rm conf}$. At fixed measured $\alpha$ and $B$, $Q$ is proportional to $\gamma_{b}$, and other published bands for La-based cuprates at this doping multiply it by $0.87$ to $2$ (Sec.~\ref{sec:Sdata}). Even the least favorable combination gives $Q=6.8$, so transport and calorimetry alone exclude a conformal bath.

The bath is also slow. The optical rate has a high-frequency slope $0.616$--$0.635$~\cite{Michon2023}, so $\lambda=0.40$, and hence $R_{\rm opt}=8.0$--$9.1$, $1.7$--$1.9$ times the conformal $4.69$ (Fig.~\ref{fig:data}b), $C_{b}=\alpha/\lambda=12.6$--$14.3$ and $\beta=3.6$--$4.1$. The optical mass logarithm of Ref.~\cite{Michon2023}, which corresponds to $\lambda=0.46$, gives instead $C_{b}=10.8$--$12.2$ and $\beta=3.05$--$3.5$ (Sec.~\ref{sec:Sdata}). Nodal photoemission, although in Bi-based cuprates, gives $\beta=3$--$4$ independently~\cite{Reber2019,Smit2021}, consistent with $R_{\rm opt}\simeq2.3\beta$. Michon et al.\ noticed the excess: with the conformal coefficient, the coupling from their mass logarithm gives $60\%$ of the measured slope~\cite{Michon2023}. The bath thus relaxes two to four times more slowly than a conformal one. This argument against the conformal bath is largely independent of the one from $Q$, because the optical $\beta$ shares only $\alpha$ with $Q$ and the photoemission $\beta$ shares nothing.

The entropy share is less decisive. With the same $\gamma_{b}$ and $B$, $\eta=\gamma_{b}\lambda/B=0.70$ ($0.62$--$0.82$) for $\lambda=0.40$ and $0.81$ ($0.72$--$0.94$) for $\lambda=0.46$. Because $\eta\propto\gamma_{b}$, it inherits the same band factor of $0.87$ to $2$, and the larger factors, from bands closer to the van Hove point, would put $\eta$ above one (Sec.~\ref{sec:Sdata}). We keep Michon et al.'s band, chosen independently of this framework, but not their normalization of the spectral weight, which assumes $b=1$. The entropy alone then places LSCO anywhere from the conformal $\frac23$ to the field's $0.93$--$0.99$ and does not decide. Given $\lambda$, $\eta=Q/C_{b}$ follows from $Q$ and $\beta$, so it adds a position on the map rather than another test.

In Nd-LSCO at $p=0.24$ ($B=2.87$ above $2$~K), the in-plane slope~\cite{Daou2009,Legros2019} gives $Q=8.3\pm1.4$, $2.5$ standard deviations above $Q_{\rm conf}$, while the isotropic ADMR rate~\cite{Grissonnanche2021}, measured on another crystal, gives $Q=4.6\pm1.6$, consistent with a conformal bath. With LSCO's $C_{b}$ the central values give $\eta=0.58$--$0.77$ and $0.32$--$0.42$, neither near the field's $0.93$; with a conformal bath they give $\eta=1.13\pm0.19$, above the local bound, and $0.62\pm0.22$, consistent with $\frac23$. The $\beta$ of Nd-LSCO is unmeasured, and the factor $1.8$ between the two routes points to a transport discrepancy between samples or methods (Sec.~\ref{sec:Sdata}).

The random couplings are themselves required. An isotropic $T$-linear rate with a momentum-independent self-energy is what zero-mean random couplings produce~\cite{PatelGuoEsterlisSachdev2023}, and within the large-$N$ family $(\psi^{\dagger}\psi)^{n}\phi^{m}$ only $n=m=1$ does so in two dimensions~\cite{SinWang2025}. The data do not fix $u$: $\lambda=0.40$ favors a finite $u$, and $\lambda=0.46$ the fixed-amplitude limit.

The field also predicts a drift. Its $C_{b}$, and with it the resistivity slope, rises on cooling by $3$--$17\%$ per decade, so $\rho-\rho_{0}\propto T^{n}$ with $n=0.956$--$0.988$ over $1$--$27$~K, $0.954$--$0.988$ over $1.4$--$30$~K and $0.934$--$0.986$ over $40$--$85$~K, whereas a conformal bath gives $n=1$. Power-law fits to published curves over these windows (Fig.~\ref{fig:field}b; Sec.~\ref{sec:Sdata}) give $n=0.955$--$0.985$ for Nd-LSCO at $33$~T, inside the predicted range, and $n=1.012\pm0.017$ for LSCO at $p=0.23$ and $48$~T, $1.4$ standard deviations above the predicted range and $2.3$ above its fixed-amplitude part. The zero-field LSCO crystal of the optics gives $n=1.03$ ($1.02$--$1.04$ across sub-windows), above both predictions. Magnetoresistance~\cite{GiraldoGallo2018}, superconducting fluctuations and the superlinear term of overdoped LSCO~\cite{Cooper2009}, which the zero-field curve shows already below $85$~K, enter at the same $1$--$5\%$ level as the predicted curvature. Taken at face value, the LSCO curves lean against the drift and the Nd-LSCO curve toward it, so the published curves do not decide.

Outside the cuprates, Sr$_{3}$Ru$_{2}$O$_{7}$ at $7.9$~T has $Q\approx4.7$ ($3.7$--$6.1$)~\cite{Tamai2008,Bruin2013,Rost2011}. A measured $\beta=2$ would place it at the conformal point, and $\beta\approx3.2$ on a slow field with $\eta\approx0.4$ (Sec.~\ref{sec:Sdata}). Four Kondo lattices give $Q=0.2$--$1.8$~\cite{TaupinPaschen2022,Custers2003,Bianchi2003,Lohneysen2007}, so $\eta\approx0.02$--$0.25$ if the framework applies: most of their critical entropy does not relax current. A uniform coupling to the critical field~\cite{PatelGuoEsterlisSachdev2023} or a reservoir that damps it, such as unscreened moments, lowers $\eta$. For a reservoir, $\beta$ moves only logarithmically and $\eta=c_{L}/(c_{L}+c_{R})$, the electrons' share of the field's damping (Sec.~\ref{sec:Sfield}).

\textit{Discussion.}---The specific-heat logarithm has been compared with the resistivity at a common coupling~\cite{Varma2020,Michon2023}, and the Planckian coefficient written as a ratio of rate to mass enhancement~\cite{Esterlis2021,PatelGuoEsterlisSachdev2023}, but each comparison still carried the coupling as an additional parameter. The two limits were also understood separately, the conformal bath through its transport coefficient and Lorenz ratio~\cite{PatelMcGreevyArovasSachdev2018,MaebashiVarma2025}, and the random-coupling field through its logarithmic thermal mass~\cite{Esterlis2021,PatelGuoEsterlisSachdev2023,SachdevReview2025}, the uniform-coupling route to $\eta<1$~\cite{PatelGuoEsterlisSachdev2023} and its agreement with key features of the LSCO optics~\cite{Li2024}. What was missing was a quantity in which the coupling cancels.

The factorization $Q=C_{b}\eta$ supplies it. Through $C_{b}\simeq3.6\beta$ and $R_{\rm opt}$ it ties transport to photoemission and optics, and through $\eta$ it asks where the critical entropy sits. The conformal bath is a single point on the $(\beta,\eta)$ map, and its $\eta=\frac23$ bounds $Q\le Q_{\rm conf}$, as the SYK solution confirms. The self-damped field is a short strip just below $\eta=1$, with no residual entropy and a floor on $\beta$, and a reservoir moves it down in $\eta$ at nearly fixed $\beta$. Each bath thus occupies its own region of the map, and published data can already be placed on it.

The decisive measurement is the resistivity, specific heat and optical conductivity of one crystal at the critical point, with its band from photoemission or quantum oscillations to fix $\gamma_{b}$. A bath conformal in time would give $\beta\approx2$ and $\eta\approx\frac23$, and the self-damped field $\beta\gtrsim2.8$ with $\eta\approx0.93$--$0.99$. A slow bath with $\eta$ clearly below $0.9$ would point to a uniform coupling or a reservoir, which quantum-oscillation masses separate, since a uniform coupling dresses the single-particle mass and a reservoir does not. A $\beta$ that is constant in temperature but differs from $2$ would signal a local fixed point that neither limit provides, and $\eta>1$ or $R_{\rm opt}\neq2.3\beta$ would show that the scattering is not local. LSCO is not conformal and lies between the field with and without such couplings, or outside the local class if its band is closer to the van Hove point than Michon et al.'s. A zero-field-extrapolated resistivity over $1$--$100$~K on one crystal at $p^{*}$, with the magnetoresistance removed point by point, would settle the drift, and the in-plane resistivity of the ADMR crystal would locate the factor $1.8$ in Nd-LSCO.

The charge response and the Lorenz ratio also separate the conformal bath from the field. Because the field couples to the density, it appears in the local charge response as a momentum-broad quasi-elastic continuum that reaches half its plateau at $\hbar\omega\approx r\kB T\approx0.45\,\kB T$, with $4/\pi r\approx2.8$ times the conformal low-frequency weight (Sec.~\ref{sec:Sfield}). This is within reach of electronic Raman, resonant inelastic x-ray and momentum-resolved electron energy-loss spectroscopies, and the symmetry of the continuum would identify the field, with B$_{1g}$ pointing to nematic fluctuations~\cite{Ishida2020,Auvray2019} and A$_{1g}$ to the hybridization field of a transition from a fractionalized to a conventional Fermi liquid~\cite{SenthilSachdevVojta2003,SachdevReview2025}. Once the elastic part is removed~\cite{TulipmanBerg2023}, the inelastic Lorenz ratio is $0.713\,L_{0}$ for the conformal bath~\cite{PatelMcGreevyArovasSachdev2018,MaebashiVarma2025} and $0.76$--$0.80\,L_{0}$ for the field (Sec.~\ref{sec:Sfield}).

The main assumptions are locality, as ADMR suggests, and that the photoemission exponent of Bi-based cuprates applies to La-based ones. The two determinations of $\lambda$ in LSCO differ by $15\%$, $\eta$ inherits the band uncertainty near the van Hove point, and the floor is a large-$M$ result (Sec.~\ref{sec:Sother}). We leave the identification of the microscopic field for future work.

Strange metals are usually compared only through $\alpha^{*}$, which is close to one in very different materials, and adding the entropy can separate them. In LSCO, $Q$ already exceeds the conformal bound and the exponent independently shows a slow bath, but the entropy share is not yet precise enough to confirm the field. Same-crystal measurements across cuprates, ruthenates and heavy fermions would turn the $(\beta,\eta)$ diagram into a classification of strange-metal baths, and each class would tell the microscopic theory what it must produce.

\begin{acknowledgments}
D.~\v{S}. and D.~S. are partially supported by the US National Science Foundation, under Grant No.~PHY-2310363. Anthropic's Claude was used in assisting with text and numerical code, as well as auditing earlier versions of the manuscript. 
\end{acknowledgments}

\bibliography{refs_v6}

\clearpage
\onecolumngrid
\setcounter{secnumdepth}{3}
\setcounter{section}{0}
\setcounter{equation}{0}
\setcounter{figure}{0}
\setcounter{table}{0}
\renewcommand{\thesection}{S\arabic{section}}
\renewcommand{\theequation}{S\arabic{equation}}
\renewcommand{\thefigure}{S\arabic{figure}}
\renewcommand{\thetable}{S\arabic{table}}
\renewcommand{\theHsection}{S\arabic{section}}
\renewcommand{\theHequation}{S\arabic{equation}}
\renewcommand{\theHfigure}{S\arabic{figure}}
\renewcommand{\theHtable}{S\arabic{table}}

\begin{center}
{\large\textbf{Supplemental Material}}
\end{center}

\section{The local marginal bath}
\label{sec:Sbath}
Electrons with density of states $N_{0}$ per spin and per site couple through the local density to a bath that is uncorrelated between sites. The self-energy is local, with Eliashberg function $\aF(\Omega,T)=-(g^{2}N_{0}/\pi)\Im D^{R}_{\rm loc}(\Omega,T)$, where $g^{2}\to\overline{g'^{2}}$ for random vertices. The coupling $\lambda$ is defined by the flat high-frequency value $\aF\to\lambda/2$. Then $1-\partial_{\varepsilon}\Re\Sigma=1+\lambda\ln(\omega_{c}/|\varepsilon|)$ and $\gamma_{b}=\frac{2\pi^{2}}{3}\kB^{2}N_{0}$. We use $\hbar=\kB=1$. The symbol $\gamma$ without subscript is the Landau-damping coefficient of the field.

The single-particle rate is $-\Im\Sigma(\varepsilon)=\pi\int_{0}^{\infty}d\Omega\,\aF(\Omega)[2n_{B}(\Omega)+f(\Omega+\varepsilon)+f(\Omega-\varepsilon)]$. At $\varepsilon=0$, $n_{B}+f=1/\sinh(\Omega/T)$, which gives $-\Im\Sigma(0,T)=\frac{\pi}{2}\lambda T\beta$ with $\beta$ of Eq.~(\ref{eq:beta}). For $\varepsilon\gg T$ the bracket tends to $\theta(\varepsilon-\Omega)$, so $-\Im\Sigma\to\frac{\pi}{2}\lambda|\varepsilon|$. For the conformal form, $-\Im\Sigma=\frac{\pi}{2}\lambda\varepsilon\coth(\varepsilon/2T)$, so $\beta=2$. For $w=(2/\pi)\arctan(x/r)$, $\beta=4.04$, $3.47$, $3.35$, $3.18$, $3.08$, $2.13$ and $1.31$ at $r=0.30$, $0.42$, $0.45$, $0.50$, $0.53$, $1$ and $2$, and $\beta=2$ at $r=1.104$. A power-law-liquid fit~\cite{Reber2019}, $\Im\Sigma\propto[\varepsilon^{2}+(\beta T)^{2}]^{a}$, returns $\beta\simeq$ Eq.~(\ref{eq:beta}) when $a\simeq\frac12$. A larger fitted $a$ (up to $0.84$ in overdoped Bi2201~\cite{Smit2021}) breaks the identification. Reference~\cite{Reber2019} compares its $\beta=3.5\pm0.5$ with the value $\pi$ of a Fermi liquid, a different reading of the same number.

Vertex corrections vanish for a momentum-independent vertex and self-energy, and $\Re\Sigma$ drops out of the dc conductivity~\cite{PatelMcGreevyArovasSachdev2018,PrangeKadanoff1964}. Hence $\sigma=e^{2}\langle Nv_{x}^{2}\rangle\langle\tau\rangle$ with $\langle\tau\rangle=\int d\varepsilon\,(-\partial_{\varepsilon}f)/2|\Im\Sigma|$, and $C_{b}\equiv1/(\lambda T\langle\tau\rangle)$. For $\tanh(x/2)$, $\langle\tau\rangle=(\pi\lambda T)^{-1}\int_{0}^{\infty}dx\,\tanh(x/2)/[2x\cosh^{2}(x/2)]=(\pi\lambda T)^{-1}\,7\zeta(3)/2\pi^{2}$, so $C_{b}=2\pi^{3}/7\zeta(3)=7.370$, in agreement with Ref.~\cite{PatelMcGreevyArovasSachdev2018} ($0.120251=7\zeta(3)/4\pi^{5/2}$) and Ref.~\cite{MaebashiVarma2025} (Eq.~61). Table~\ref{tab:shapes} lists other forms. Over six families of shapes [$\tanh(x/2s)$, $1-e^{-x/s}$, $\min(1,x/s)$, $x^{2}/(s^{2}+x^{2})$, $\tanh^{2}(x/2s)$, ${\rm erf}(x/s)$, each at six widths $s=0.5$--$3$], the $24$ shapes with $\beta\ge1.5$ have $C_{b}/\beta=3.48$--$3.76$; the arctan family over $r=0.3$--$2$ has $3.49$--$3.71$.

\begin{table}[htb]
\caption{$\beta$ and $C_{b}$ for several thermal forms $w(x)$, and the optical ratio $R_{\rm opt}=2C_{b}/\pi$.}
\label{tab:shapes}
\begin{ruledtabular}
\begin{tabular}{lccc}
$w(x)$ & $\beta$ & $C_{b}$ & $R_{\rm opt}$ \\ \hline
$\tanh(x/2)$ (conformal) & $2.00$ & $7.37$ & $4.69$ \\
$\frac{2}{\pi}\arctan(x/1.104)$ & $2.00$ & $7.28$ & $4.64$ \\
$1-e^{-x}$ & $2.77$ & $9.98$ & $6.35$ \\
$\tanh x$ & $3.14$ & $11.26$ & $7.17$ \\
$\min(1,x)$ & $3.44$ & $12.28$ & $7.82$ \\
$\frac{2}{\pi}\arctan(x/0.45)$ & $3.35$ & $11.84$ & $7.54$ \\
$\frac{2}{\pi}\arctan(x/0.30)$ & $4.04$ & $14.13$ & $9.00$ \\
\end{tabular}
\end{ruledtabular}
\end{table}

The optical ratio follows from the extended-Drude form. In the convention $\sigma=iK/(\omega+M)$, $M=\omega(m^{*}/m-1)+i\hbar/\tau$, a local self-energy gives $\hbar/\tau(0)=C_{b}\lambda T$ and $\hbar/\tau\to\frac{\pi}{2}\lambda\omega$ for $\omega\gg T$. We checked both by evaluating $\sigma(\omega)=(i/\omega)\int d\varepsilon\,[f(\varepsilon)-f(\varepsilon+\omega)]/[\omega+\Sigma^{*}(\varepsilon)-\Sigma(\varepsilon+\omega)]$ with the full Kramers--Kronig $\Re\Sigma$. Rescaling $K$ multiplies both by the same factor, so $R_{\rm opt}$ does not depend on $K$ or $\lambda$. Beyond the zero-frequency value, the conformal and arctan forms have nearly the same $\omega$ dependence at fixed $\lambda$ (Fig.~\ref{fig:data}b).

The entropy of a bath with dynamics of its own follows from the free energy. To order $g^{2}$ this is the bubble joining the local density correlator to the bath. For a constant density of states $\Im\Pi_{\rm loc}=2\pi N_{0}^{2}\Omega$ exactly, and
\begin{equation}
F_{2}(T)=-\frac{3\gamma_{b}}{2\pi^{2}}\int_{0}^{\infty}d\Omega\,\Omega\coth\frac{\Omega}{2T}\,\Lambda(\Omega,T),\qquad \Lambda(\Omega,T)=2\,{\rm P}\!\int_{0}^{\infty}d\omega\,\frac{\omega\,\aF(\omega,T)}{\omega^{2}-\Omega^{2}} .
\label{eq:F2}
\end{equation}
A flat spectrum gives $\delta(C/T)=\gamma_{b}\lambda\ln(\omega_{c}/T)$, i.e.\ $b=1$. If weight is removed below $T$, Kramers--Kronig shifts $\Lambda$ at $\Omega\gg T$ by $\lambda T^{2}I_{w}/\Omega^{2}$ with
\begin{equation}
I_{w}=\int_{0}^{\infty}dx\,x\,[1-w(x)],
\end{equation}
which adds $3I_{w}/\pi^{2}$ to $b$. For $\tanh(x/2)$, $I_{w}=\pi^{2}/6$ and $b=\frac32$. Direct evaluation with the digamma kernel gives log slopes $9.873$ and $6.579$, against $\pi^{2}$ and $2\pi^{2}/3$. With $G_{\rm loc}$ independent of $\Sigma$ (flat band, $\omega_{c}\ll W$), the $\Sigma$-linear part of ${\rm Tr}\ln(-G^{-1})$ cancels ${\rm Tr}\,\Sigma G$, so $F=F_{0}-\Phi$ is exactly linear in $\lambda$, and $b=\frac32$ holds to all orders. This result requires a convergent $I_{w}$ and bath dynamics that do not depend on the electrons, as for SYK islands or an AdS$_{2}$ sector; the two-band SYK solution confirms it to $0.2\%$.

\begin{figure}[htb]
\includegraphics[width=\columnwidth]{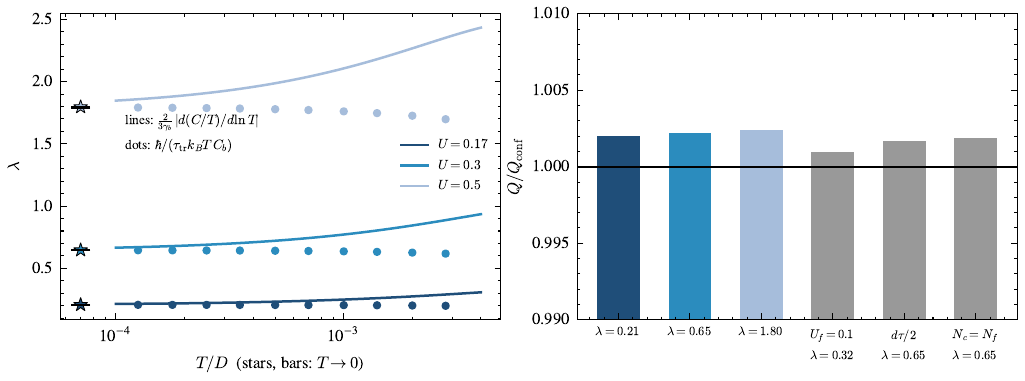}
\caption{Two-band SYK model~\cite{Chowdhury2018}, $U_{f}=0.05D$. Left: the specific-heat coupling $\lambda_{C}=\frac{2}{3\gamma_{b}}|d(C/T)/d\ln T|$ (lines) and the transport coupling $\lambda_{\rm tr}=\hbar/(\ttr\kB TC_{b})$ (dots) for $U=0.17$, $0.3$, $0.5$. Right: $Q/Q_{\rm conf}=\lambda_{\rm tr}/\lambda_{C}$ at $T\to0$ for the frozen-bath runs, for $U_{f}=0.1D$, for half the time step, and with bath feedback at $N_{c}=N_{f}$.}
\label{fig:syk}
\end{figure}

For the two-band SYK check we solve $\Sigma_{c}=U^{2}G_{c}G_{f}^{2}$ and $\Sigma_{f}=U_{f}^{2}G_{f}^{3}+r_{N}U^{2}G_{f}G_{c}^{2}$ ($r_{N}=N_{c}/N_{f}$) at half filling with a semicircular band of half-width $D=1$~\cite{Chowdhury2018}. The free energy is computed on the Matsubara axis (Luttinger--Ward form, fit $\Delta F=a_{0}+a_{2}T^{2}+\frac{B}{2}T^{2}\ln T+\dots$ over $10^{-4}\le T\le4\times10^{-3}$, residual $10^{-12}$). The Kubo rate is computed on the real axis ($2^{21}$ points). The two solvers agree on $\Sigma_{c}(i\omega_{n})$ to $10^{-4}$. Over a ninefold range of $\lambda$ ($0.21$--$1.80$), $Q/Q_{\rm conf}=1.0020$, $1.0022$ and $1.0024$; doubling $U_{f}$ reduces the excess to $1.0010$, identifying it as the finite-$T/U_{f}$ extrapolation. Halving the time step changes $Q$ by $0.05\%$, and feedback at $N_{c}=N_{f}$ by $0.03\%$ (Fig.~\ref{fig:syk}).

\section{The self-damped field}
\label{sec:Sfield}
Following Refs.~\cite{Esterlis2021,PatelGuoEsterlisSachdev2023}, an $M$-component field couples to the local electron density through spatially random vertices of zero mean and variance $\overline{g'^{2}}$. At large $M$ the disorder-averaged field has, per component,
\begin{equation}
D^{-1}(q,i\Omega_{n})=m+\varepsilon_{q}+\gamma|\Omega_{n}|,\qquad \varepsilon_{q}=\nu_{0}q^{2}\in[0,E_{L}],\qquad \gamma=2\pi\overline{g'^{2}}N_{0}^{2},
\label{eq:D}
\end{equation}
where $\gamma|\Omega_{n}|$ is the Landau damping by the electrons ($\Im\Pi_{\rm loc}=2\pi N_{0}^{2}\Omega$) and $\rho=1/4\pi\nu_{0}$ is the mode density per unit $\varepsilon$. The local spectral function is $-\Im D^{R}_{\rm loc}=\rho[\arctan(\gamma\Omega/m)-\arctan(\gamma\Omega/(m+E_{L}))]$, so
\begin{equation}
\aF=\frac{\lambda}{2}\,\frac{2}{\pi}\Big[\arctan\frac{\gamma\Omega}{m}-\arctan\frac{\gamma\Omega}{m+E_{L}}\Big],\qquad \lambda=M\,\overline{g'^{2}}N_{0}\rho ,
\label{eq:a2Ffield}
\end{equation}
which is the arctan form $w=(2/\pi)\arctan(x/r)$ with $r=m/\gamma T$ for $\gamma\Omega\ll E_{L}$.

For a local self-energy and a flat band, the fermion determinant together with the Luttinger--Ward functional gives the free band, $\gamma_{b}$, and the whole enhancement of $C/T$ comes from the field (Ref.~\cite{Esterlis2021}, Sec.~VIII and App.~B). The field part at the large-$M$ saddle is
\begin{equation}
F_{\phi}(m,T)=\frac{M}{2}\,T\sum_{n}\rho\int_{0}^{E_{L}}d\varepsilon\,\ln\!\big(m+\varepsilon+\gamma|\Omega_{n}|\big)-\frac{M}{4u}\,(m-m_{c})^{2},
\label{eq:Fphi}
\end{equation}
where $u$ is the self-interaction and $m_{c}$ is the bare mass that puts the $T=0$ system at its critical point. Stationarity, $\partial F_{\phi}/\partial m=0$, is the saddle equation $m/u=\langle\phi^{2}\rangle_{T}(m)-\langle\phi^{2}\rangle_{0}(0)$ solved below; it fixes $m_{c}=-u\langle\phi^{2}\rangle_{0}(0)$.

Because $\partial F_{\phi}/\partial m=0$ at the saddle, $S=-dF/dT=-\partial F/\partial T|_{m}$. The second term of Eq.~(\ref{eq:Fphi}) has no explicit $T$ dependence. For the first we use the phase representation of a damped mode: with $\varphi_{\varepsilon}(\Omega)=\arctan[\gamma\Omega/(m+\varepsilon)]$ the phase of $D^{-1}_{R}$, $\frac12T\sum_{n}\ln(m+\varepsilon+\gamma|\Omega_{n}|)=\int_{0}^{\infty}\frac{d\Omega}{\pi}\,\varphi_{\varepsilon}'(\Omega)\,[\frac{\Omega}{2}+T\ln(1-e^{-\Omega/T})]+{\rm const}$. Hence
\begin{equation}
S=S_{\rm band}+M\int_{0}^{\infty}d\Omega\,g(\Omega)\,\sigma(\Omega/T),\qquad g(\Omega)=\frac{\rho\gamma}{2\pi}\ln\frac{(m+E_{L})^{2}+\gamma^{2}\Omega^{2}}{m^{2}+\gamma^{2}\Omega^{2}},
\label{eq:SMSfield}
\end{equation}
with $\sigma(x)=x/(e^{x}-1)-\ln(1-e^{-x})$ the entropy of one oscillator. For $\gamma T\ll E_{L}$ and $m=r\gamma T$, $g(xT)=(\rho\gamma/\pi)[\ln(E_{L}/\gamma T)-\frac12\ln(r^{2}+x^{2})]$, and with $\int_{0}^{\infty}\sigma(x)\,dx=\pi^{2}/3$,
\begin{equation}
S-S_{\rm band}=\frac{M\rho\gamma T}{\pi}\Big[\frac{\pi^{2}}{3}\ln\frac{E_{L}}{\gamma T}-K(r)\Big],\qquad K(r)=\frac12\int_{0}^{\infty}dx\,\sigma(x)\ln(r^{2}+x^{2}).
\label{eq:Slog}
\end{equation}
This vanishes as $T\ln T$ for $T\to0$: there is no residual entropy. At constant $r$, $C/T=dS/dT$ has the logarithmic coefficient
\begin{equation}
B=\frac{\pi M\rho\gamma}{3}=\frac{2\pi^{2}}{3}\,M\,\overline{g'^{2}}N_{0}^{2}\rho=\gamma_{b}\lambda ,
\label{eq:identity}
\end{equation}
by Eq.~(\ref{eq:a2Ffield}) and $\gamma_{b}=\frac{2\pi^{2}}{3}N_{0}$. So $b=1$ at leading logarithm: the damping coefficient $\gamma$ that sets the field's entropy is the one that sets $\lambda$, because both come from the same electron--field vertex. The temperature dependence of $r$ adds $-(M\rho\gamma/\pi)\,d[TK(r)]/dT$ to $C$, whose logarithmic part is of relative order $d\ln r/d\ln T\sim1/\ln(E_{L}/T)$; the saddle solved below gives $\delta\equiv b-1=0.007$--$0.074$ for $\gamma T/E_{L}=10^{-5}$--$10^{-2}$, i.e.\ $\eta=1/(1+\delta)=0.93$--$0.99$.

The fixed-bath formula would instead produce a constant entropy, and the way it does shows why the field has none. Inserted into Eq.~(\ref{eq:F2}), the arctan form has $1-w\simeq2r/\pi x$ at large $x$ and $I_{w}\simeq(2r/\pi)\,\omega_{c}/T$ diverges at the cutoff. The divergent piece is a $T$-linear free energy, i.e.\ a temperature-independent entropy that would survive to $T=0$. In Eq.~(\ref{eq:Fphi}) this piece is the term linear in $m$, $\frac{M}{2}m\langle\phi^{2}\rangle_{0}$, with an ultraviolet $\langle\phi^{2}\rangle_{0}$; with $m=r\gamma T$ it is linear in $T$. The critical tuning $m_{c}=-u\langle\phi^{2}\rangle_{0}(0)$ cancels it exactly, which is the statement that the system sits at its critical point, and stationarity guarantees that the remaining $m(T)$ dependence enters $S$ only through Eq.~(\ref{eq:SMSfield}), which is finite and vanishes as $T\to0$. The fixed-bath formula fails here because the field's dynamics are made by the electrons: changing the coupling changes $\gamma$ and $\lambda$ together. The conformal bath of Sec.~\ref{sec:Sbath} has dynamics of its own, a convergent $I_{w}$, and $b=\frac32$.

The thermal mass follows from the saddle equation. In units $\gamma=E_{L}=1$ with frequency cutoff $E_{c}$, the saddle equation is $m/u=\langle\phi^{2}\rangle_{T}(m)-\langle\phi^{2}\rangle_{0}(0)$ with $\langle\phi^{2}\rangle_{0}(m)=(\rho/\pi)\int_{0}^{E_{L}}d\varepsilon\,\ln[(m+\varepsilon+E_{c})/(m+\varepsilon)]$ and the thermal part $(\rho/\pi)\int_{0}^{\infty}d\Omega\,2n_{B}(\Omega)[\arctan((m+E_{L})/\Omega)-\arctan(m/\Omega)]$. We evaluate $S$ from Eq.~(\ref{eq:SMSfield}) along $m(T)$, then $C/T=dS/dT$, $B=-d(C/T)/d\ln T$ and $\delta=B/B_{0}-1$ with $B_{0}=\pi\rho/3$. Table~\ref{tab:saddle} gives $r=m/T$: it increases with $u$ and saturates. At $u\to\infty$ the saddle becomes the constraint $\langle\phi^{2}\rangle_{T}(m)=\langle\phi^{2}\rangle_{0}(0)$ of a fixed-amplitude field, the analogue of the spherical constraint, and $u$ drops out. Numerically this limit is taken at $u\rho=10^{3}$. Table~\ref{tab:floor} gives the resulting $\beta$, $C_{b}$, $\delta$ and $Q$. Across $E_{c}=0.3$--$3$, $u\rho\ge0.3$ and $\gamma T/E_{L}=10^{-5}$--$10^{-2}$: $\beta=2.8$--$4.5$, $\delta=0.007$--$0.074$, $Q=9.2$--$15.5$; in the fixed-amplitude limit $\beta\ge2.8$. In the model of Refs.~\cite{Esterlis2021,Li2024} the fixed-amplitude limit has the closed form $r=\pi W_{0}(\ell/\pi)/\ell$ with $\ell=\ln(a_{0}e^{\gamma_{E}}/2\pi T)$ (Eq.~(8.20) of Ref.~\cite{Esterlis2021}); a finite $u$ adds a positive term $\propto1/u$ to $\ell$, which lowers $r$. That form gives $r=0.45$ at $\ell=5.57$, i.e.\ $a_{0}/T\approx9\times10^{2}$.

\begin{table}[htb]
\caption{$r=m/(\gamma T)$ from the saddle at $E_{c}=E_{L}$ (the last row is $u\rho=10^{3}$).}
\label{tab:saddle}
\begin{ruledtabular}
\begin{tabular}{lcccc}
$u\rho$ $\backslash$ $\gamma T/E_{L}$ & $10^{-5}$ & $10^{-4}$ & $10^{-3}$ & $10^{-2}$ \\ \hline
$0.3$ & $0.230$ & $0.248$ & $0.268$ & $0.293$ \\
$1$ & $0.299$ & $0.333$ & $0.379$ & $0.442$ \\
$3$ & $0.330$ & $0.374$ & $0.437$ & $0.534$ \\
$10$ & $0.342$ & $0.392$ & $0.463$ & $0.580$ \\
$\infty$ & $0.348$ & $0.400$ & $0.476$ & $0.603$ \\
\end{tabular}
\end{ruledtabular}
\end{table}

\begin{table}[htb]
\caption{The fixed-amplitude limit ($u\rho=10^{3}$, $E_{c}=E_{L}$): $r$, $\beta$, its drift, $C_{b}$, $\delta=b-1$ and $Q=C_{b}/(1+\delta)$.}
\label{tab:floor}
\begin{ruledtabular}
\begin{tabular}{lcccccc}
$\gamma T/E_{L}$ & $r$ & $\beta$ & $d\beta/d\log_{10}T$ & $C_{b}$ & $\delta$ & $Q$ \\ \hline
$7.1\times10^{-5}$ & $0.390$ & $3.59$ & $-0.25$ & $12.6$ & $0.028$ & $12.3$ \\
$2.2\times10^{-4}$ & $0.423$ & $3.45$ & $-0.28$ & $12.2$ & $0.033$ & $11.8$ \\
$7.1\times10^{-4}$ & $0.462$ & $3.31$ & $-0.32$ & $11.7$ & $0.040$ & $11.2$ \\
$2.3\times10^{-3}$ & $0.513$ & $3.13$ & $-0.36$ & $11.1$ & $0.049$ & $10.6$ \\
$7.2\times10^{-3}$ & $0.580$ & $2.94$ & $-0.42$ & $10.5$ & $0.056$ & $9.9$ \\
\end{tabular}
\end{ruledtabular}
\end{table}

A reservoir that also damps the field lowers $\eta$. Let the field be damped by the electrons, $c_{L}=\gamma=2\pi\overline{g'^{2}}N_{0}^{2}$, and by an Ohmic reservoir, $c_{R}$, so that $\gamma\to c_{L}+c_{R}$ in Eq.~(\ref{eq:D}). The plateau of Eq.~(\ref{eq:a2Ffield}), which sets $\lambda$ and the rate, does not depend on the damping, so $\gamma_{b}\lambda\propto c_{L}$, while Eq.~(\ref{eq:identity}) gives $B\propto c_{L}+c_{R}$. Hence $\eta=c_{L}/(c_{L}+c_{R})$ at leading logarithm, if the reservoir carries no logarithm of its own (if it does, $\eta$ is smaller still). The bandwidth $E_{L}$ shrinks as $\eta E_{L0}$, which enters $r$ only through the logarithm: in the fixed-amplitude limit at $T/E_{L0}=5\times10^{-4}$, $\beta=3.35$, $3.26$, $3.00$ and $2.69$ at $\eta=1$, $0.5$, $0.1$ and $0.02$. A reservoir therefore moves a material down in $\eta$ at nearly fixed $\beta$; it does not reach the conformal point.

A uniform coupling to the field is the other route to $\eta<1$, and the one usually considered for heavy fermions: following Ref.~\cite{PatelGuoEsterlisSachdev2023}, Eq.~(13), $b-1\simeq(g^{2}/\Gamma Ng'^{2})(L_{2}/L_{1})$. Related settings for the reservoir route, a hybridization boson damped by two fermion species~\cite{Aldape2022} and an extra damping channel that enhances the boson specific heat~\cite{Caprara2022}, have been studied, but we have not found the ratio $\eta=c_{L}/(c_{L}+c_{R})$ stated. A quantum-oscillation or photoemission mass that accounts for $C/T$ would indicate the uniform-coupling route; a mass that falls short, with $b\approx1/\eta$ from optics, a reservoir.

\begin{figure}[htb]
\includegraphics[width=\columnwidth]{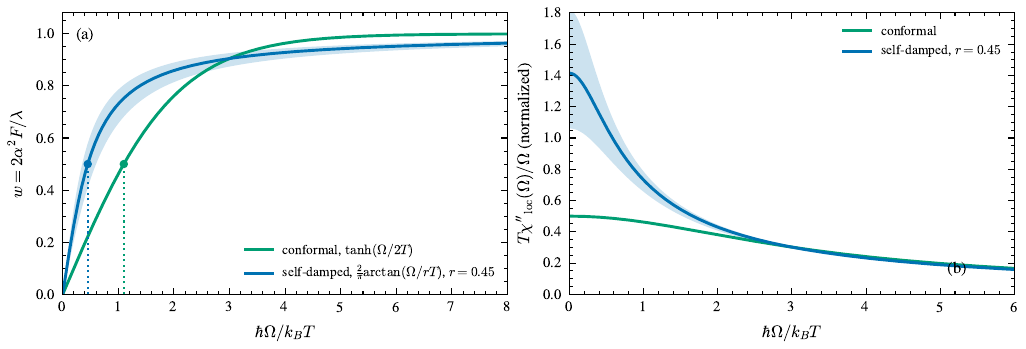}
\caption{(a)~Local bath spectra $w=2\aF/\lambda$: conformal, $\tanh(\Omega/2T)$ (green), and the overdamped two-dimensional field, $(2/\pi)\arctan(\Omega/r\kB T)$ with $r=0.45$ (blue; shading $r=0.35$--$0.6$). Dots: the frequencies at which each reaches half strength, $0.45$ and $1.10\,\kB T$. (b)~Local quasi-elastic charge weight $T\chi''_{\rm loc}(\Omega)/\Omega$, normalized to the flat plateau, for the same spectra.}
\label{fig:spectra}
\end{figure}

The field also leaves signatures of its own. Because the field couples to the local density, it appears in the local charge response as a momentum-broad relaxational continuum. $\chi''_{\rm loc}$ reaches half its plateau at $\hbar\omega\approx r\kB T\approx0.45\,\kB T$; $\chi''_{\rm loc}/\omega$ at low frequency is $4/\pi r\approx2.8$ times the conformal value and falls to half at $\hbar\omega\approx2.33\,r\kB T\approx1.05\,\kB T$ (Fig.~\ref{fig:spectra}). Existing electron energy-loss continua lie at $0.1$--$2$~eV~\cite{Mitrano2018,Husain2019}, and charge-order fluctuations in YBa$_{2}$Cu$_{3}$O$_{7-\delta}$-family films relax at $0.65$--$0.9\,\kB T$~\cite{Arpaia2019,Arpaia2023}. A B$_{1g}$ signal, which elastoresistance would show as a divergent nematic susceptibility~\cite{Ishida2020,Auvray2019}, would point to nematic order coupled to the local density through random strain, with a uniform $d$-wave coupling~\cite{LedererSchattnerBergKivelson2015} contributing to $1-\eta$. An A$_{1g}$ signal would point to the hybridization field of the transition from a fractionalized to a conventional Fermi liquid~\cite{SenthilSachdevVojta2003,SachdevLectures2024,SachdevReview2025}.

The Lorenz ratio differs between the two limits as well. For purely inelastic scattering it is $0.713\,L_{0}$ for a conformal bath~\cite{PatelMcGreevyArovasSachdev2018,MaebashiVarma2025} and $0.76$--$0.80\,L_{0}$ for the self-damped field ($r=0.6$--$0.3$); the arctan form at the conformal $\beta=2$ ($r=1.10$) gives $0.731\,L_{0}$. Elastic scattering raises all of these ($0.82\,L_{0}$ for the conformal bath at $\rho_{0}/\rho=0.5$), so the test needs the inelastic part extracted from the temperature dependence~\cite{TulipmanBerg2023}.

\section{Published data}
\label{sec:Sdata}
For LSCO at $p=0.24$ we extracted the curves of Figs.~2a and 2b of Ref.~\cite{Michon2023} from the vector paths of the arXiv source file \texttt{fig02.pdf}, calibrating the axes by the frame lines and major tick marks (script \texttt{extract\_michon2023\_fig2.py}). The check is the dashed guide line of Fig.~2a, which comes out as $\rho=13.5+0.629T$~$\mu\Omega$\,cm, against the quoted slope $0.63$. An earlier extraction, whose code was lost, had placed the zero of Fig.~2b at the center of the ``0'' label, $2.2$~pt below the axis; all its rates were $0.91$ too high. Slopes are unaffected, and no number in the Letter used those intercepts.

The optical slope fixes $\lambda$ directly. Linear fits over $\hbar\omega/\kB T=20$--$80$ to the $50$--$100$~K curves give slopes $0.616$--$0.635$ (mean $0.625$, so $\lambda=2\times0.625/\pi=0.40$). The zero-frequency value is $\hbar/[\tau(0)\kB T]\approx5$ from the optics, the value $f_{\tau}(0)\approx5$ quoted in the text of Ref.~\cite{Michon2023} (our corrected extraction of their Fig.~2b gives $4.4$--$5.9$ between $100$ and $300$~K, without a trend), or $5.7$ from the measured $A=0.63\,\mu\Omega$\,cm/K with their conversion ($5\leftrightarrow A=0.55$). Hence $R_{\rm opt}=8.0$--$9.1$, $1.7$--$1.9$ times the conformal $4.69$, so $C_{b}=12.6$--$14.3$, $r=0.29$--$0.40$ and $\beta=3.6$--$4.1$. The main systematic is $\varepsilon_{\infty}$, which controls the curvature above $0.4$~eV and hence the fitted slope. The collapse is imperfect below $\hbar\omega/\kB T\approx10$ at $50$--$75$~K.

The optical mass logarithm gives an independent estimate of the shape. The ratio of the dc rate to the scale it sets is $1.47$ (optics) to $1.66$ (dc) times the conformal value: Michon et al.'s $f_{\tau}(0)\approx5$ against the conformal $C_{b}\lambda=3.40$ at their $g=0.23$, and their measured $A=0.63$ against their model's $0.38$~$\mu\Omega$\,cm/K, all quoted in the text of Ref.~\cite{Michon2023}. This gives $C_{b}=10.8$--$12.2$, $r=0.42$--$0.54$ and $\beta=3.05$--$3.47$, so that the two routes together span $C_{b}=10.8$--$14.3$ and $\beta=3.05$--$4.1$.

With the tight-binding band of Ref.~\cite{Michon2023} ($t'/t=-0.17$, $m_{b}=2.76\,m_{e}$, $\gamma_{b}=3.89$~mJ\,K$^{-2}$\,mol$^{-1}$), $\alpha=5$--$5.7$ and $B=2.2\pm0.3$~\cite{Girod2021}, $Q=\alpha\gamma_{b}/B=8.8$--$10.1$ ($7.8$--$11.7$ with the error on $B$). Directly, $\eta=\gamma_{b}\lambda/B=0.70$ ($0.62$--$0.82$) with $\lambda=0.40$ from the optical slope, and $0.81$ ($0.72$--$0.94$) with $\lambda=0.46$, twice Michon et al.'s $g=0.23$ from the optical mass logarithm. As a check, $Q/C_{b}=0.62$--$0.93$. $Q$, $\beta$ and $\eta$ are not independent measurements, since given $\lambda$ $Q=C_{b}(\beta)\,\eta$ is fixed by the other two. $Q$ alone, however, tests the conformal bath, without $\lambda$ or spectroscopy: a conformal bath has $C_{b}=7.37$ and $\eta\le\frac23$, so $Q\le4.91$, while LSCO's lowest value, $7.8$, multiplied by the least favorable band factor below, $0.87$, is still $6.8$.

The entropy share depends on the band: $\eta$ is proportional to $\gamma_{b}$ at fixed measured $\lambda$ and $B$ (or to the band's $(m/m_{e})K$ if the optical spectral weight is taken from the band). Recomputing both for other LSCO-type bands at $n=0.76$ (\texttt{band\_dos.py}) multiplies $\eta$ by $0.87$--$0.90$ for $t'/t=-0.12$, $1.10$--$1.12$ for $t'/t=-0.20$, $1.51$--$1.55$ for $t'/t=-0.25$, and $1.60$--$1.74$ for a band with $t''=-t'/2$ fitted to photoemission~\cite{Yoshida2006}, which places $p=0.24$ closer to the van Hove point; the band fitted to the ADMR of Nd-LSCO at $p=0.24$~\cite{Grissonnanche2021} gives $1.58$--$1.70$, or $2.0$ with its interlayer hopping. The band used is Michon et al.'s, which they chose independently of this framework. Bands near the van Hove point push $\eta$ above $1$, which a local bath cannot reach; we report this as a systematic rather than use it to select the band. At fixed measured $\alpha$ and $B$, the same factor multiplies $Q$. In Michon et al.'s own normalization, which imposes equal $\ln T$ coefficients for the optical mass and $C/T$ and so assumes $b=1$, their band gives $(m/m_{e})K=582$~meV against the required $583$~meV~\cite{Michon2023}.

The specific heats of Eu-LSCO and Nd-LSCO at $p=0.24$ are the vector-graphics curves of Figs.~2d and 3d of Ref.~\cite{Michon2019}, from which we extracted about $3{,}700$ points each, with axes calibrated from the tick labels (check: the reference lines come out at $4.14$ and $3.62$~mJ\,K$^{-2}$\,mol$^{-1}$, against $4.2$ and $3.6$ quoted). Fitting $C_{\rm el}/T=B\ln(T_{0}/T)$ gives, for Eu-LSCO and Nd-LSCO respectively, $B=2.63$ and $3.02$ over $0.55$--$10$~K, $2.70$ and $2.87$ over $2$--$10$~K (Nd-LSCO at $18$~T, where its Nd Schottky term is negligible), and $2.56$ and $3.72$ over $0.55$--$2$~K (Nd-LSCO at $8$~T). The Nd-LSCO excess is confined below $2$~K (about $1.6$, $1.1$ and $0.5$~mJ\,K$^{-2}$\,mol$^{-1}$ at $0.56$, $1.0$ and $1.7$~K) and is closer to $1/T$ than to a Schottky tail; it could be a residual of the Nd moments not removed by the $p=0.12$ reference subtraction, or a small doping mismatch. For Eu-LSCO over $0.55$--$10$~K the logarithm has r.m.s.\ residual $0.15$~mJ\,K$^{-2}$\,mol$^{-1}$ and the best power law, $T^{-0.17}$, has $0.23$.

For Nd-LSCO, the ADMR fit~\cite{Grissonnanche2021} gives an isotropic $T$-linear rate $\alpha=1.2\pm0.4$ with the ADMR band ($\gamma_{b}=11.0$~mJ\,K$^{-2}$\,mol$^{-1}$); with $B=2.87$, $Q=4.6\pm1.6$. The same scattering model in the zero-field Boltzmann equation gives $\Asq=4.2$~$\Omega$/K, against $7.4\pm0.8$~$\Omega$/K measured on another crystal~\cite{Daou2009,Legros2019}, which gives $Q=8.3\pm1.4$, $2.5$ standard deviations above $Q_{\rm conf}=4.91$; the ADMR value is consistent with $Q_{\rm conf}$. If Nd-LSCO has LSCO's $C_{b}=10.8$--$14.3$, the central values give $\eta=0.58$--$0.77$ (in-plane) and $0.32$--$0.42$ (ADMR); neither is near the field's $0.93$--$0.99$. If instead its bath were conformal ($C_{b}=7.37$), $\eta=1.13\pm0.19$ (in-plane, above the local bound) and $0.62\pm0.22$ (ADMR, consistent with $\frac23$). Its exponent $\beta$ has not been measured, so Nd-LSCO is not placed on the map. Grissonnanche et al.\ matched their computed $\rho_{xx}$ at $33$~T to the data with a constant factor of $0.8$~\cite{Grissonnanche2021}, so part of the factor $1.8$ between the two routes may come from comparing different samples and fields. The in-plane resistivity of the ADMR crystal would settle it.

For Sr$_{3}$Ru$_{2}$O$_{7}$, if the renormalization is local, $\omega_{p}^{2}\gamma$ is invariant, and the pockets and masses of Ref.~\cite{Tamai2008}, which reproduce $\gamma=127$~mJ\,K$^{-2}$\,mol-Ru$^{-1}$, give the band product directly. With the slope at $7.9$~T~\cite{Bruin2013} and $B$ from Ref.~\cite{Rost2011}, $Q=4.7$ ($3.7$--$6.1$). Quantum oscillations follow five sheets toward the critical field; in our reading of Ref.~\cite{Mercure2010} their masses change by at most $4\,m_{e}$ while $\gamma$ rises two- to threefold, so they account for at most about $60\%$ of it. A measured $\beta$ of $2.0$, $3.2$ or $1.3$ would place it at the conformal point ($\eta\approx\frac23$), on the reservoir line of the field ($\eta\approx0.4$), or on a fast field ($r\approx2$).

\begin{figure}[htb]
\includegraphics[width=0.55\columnwidth]{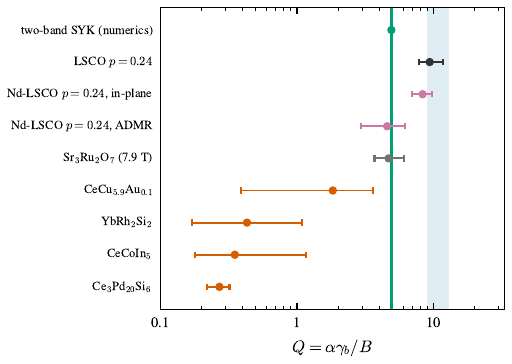}
\caption{$Q=\alpha\gamma_{b}/B$ for the systems discussed in the text. Green line: $Q_{\rm conf}=4.91$. Blue band: the self-damped field near its fixed-amplitude limit ($Q\approx9$--$13$). Nd-LSCO: the two transport routes with $B=2.87$. The SYK point is a numerical check, not a material.}
\label{fig:materials}
\end{figure}

For the Kondo lattices we obtain $Q$ from the slope, $B$ and a carrier density $n$: CeCoIn$_{5}$ $0.35$ ($0.18$--$1.17$, $n=10$--$12.5$~nm$^{-3}$); YbRh$_{2}$Si$_{2}$ $0.43$ ($0.17$--$1.09$, $n=5$--$20$~nm$^{-3}$); CeCu$_{5.9}$Au$_{0.1}$ $1.8$ ($0.4$--$3.6$, Ce density assumed, since the Hall coefficient does not saturate); Ce$_{3}$Pd$_{20}$Si$_{6}$ $0.22$--$0.32$. For $C_{b}$ between $7.37$ and $12.5$ the central values give $\eta\approx0.02$--$0.25$ (Fig.~\ref{fig:materials}); the upper ends of the ranges reach $\eta\approx0.5$ for CeCu$_{5.9}$Au$_{0.1}$. For CeCoIn$_{5}$, terahertz data~\cite{Pracht2026} with $B$ at $5$~T and $\gamma_{b}=3.3$~mJ\,K$^{-2}$\,mol$^{-1}$ give $b\approx17$ ($13$--$29$) in our working estimate, but $2$--$7$ under other input choices; a measurement of $\gamma$ in LaCoIn$_{5}$ would fix it.

\begin{figure}[htb]
\includegraphics[width=\columnwidth]{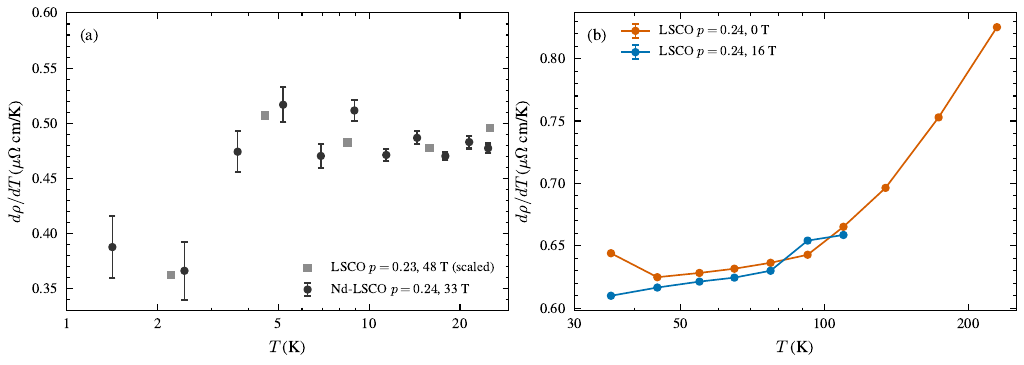}
\caption{Local resistivity slopes. (a)~Nd-LSCO at $p=0.24$ and $33$~T (dots, fits over $T$ windows, with fit errors), and LSCO at $p=0.23$ and $48$~T (squares, scaled to the same mean). (b)~LSCO at $p=0.24$, the crystal used for the optics, at $0$ and $16$~T. The self-damped field predicts a slope that rises on cooling by $3$--$17\%$ per decade; the conformal bath predicts a constant slope.}
\label{fig:drift}
\end{figure}

The self-damped field also predicts a drift of the resistivity slope. In the field, $r$ falls logarithmically on cooling (Table~\ref{tab:floor}), so $\beta$ rises by $0.12$--$0.49$ per decade of temperature and $C_{b}$, hence the transport slope, by $3$--$17\%$ per decade, depending on $u$ and $\gamma T/E_{L}$. With $\rho-\rho_{0}\propto TC_{b}(T)$ the local exponent is $n=d\ln(\rho-\rho_{0})/d\ln T=1+d\ln C_{b}/d\ln T<1$. For $E_{L}/\gamma=1$--$10$~eV the predicted power-law exponent is $n=0.956$--$0.988$ over $1$--$27$~K and $0.934$--$0.986$ over $40$--$85$~K. A conformal bath gives $n=1$. The optical intercepts cannot test this: between $100$ and $300$~K they scatter by $\pm15\%$ without a trend, while the predicted change of $C_{b}$ is $2$--$7\%$. We therefore fitted $\rho=\rho_{0}+AT^{n}$ to published resistivity curves (\texttt{drift\_test\_v6.py}; Fig.~\ref{fig:drift}).

For Nd-LSCO at $p=0.24$ and $33$~T over $1$--$27$~K~\cite{Daou2009}, extracted from the vector graphics of Fig.~1c of Ref.~\cite{Legros2019} ($3{,}246$ points), we find $n=0.955$--$0.985$, depending on the fit window (lower bound $1$--$3$~K, upper bound $20$--$27$~K), with a block-bootstrap spread of $0.007$, inside the predicted range. The local slope, however, is not monotonic: it is $0.47$~$\mu\Omega$\,cm/K at $12$--$27$~K, $0.50$ at $4$--$10$~K and $0.43$ below $3$~K, so the curvature is not clearly the predicted logarithm.

LSCO at $p=0.23$ and $48$~T over $1.4$--$30$~K~\cite{Cooper2009}, with $16$ points from Fig.~1b of Ref.~\cite{Legros2019}, gives $n=1.012\pm0.017$, consistent with $n=1$. The prediction for this window is $0.954$--$0.988$, so the fit lies $1.4$ standard deviations above the top of the range and $2.3$ above its fixed-amplitude part ($0.954$--$0.972$).

For LSCO at $p=0.24$ in zero field, the crystal of the optics (Fig.~2a of Ref.~\cite{Michon2023}), the fit over the predicted window, $40$--$85$~K, gives $n=1.031\pm0.002$, and $1.022$--$1.039$ across its sub-windows ($40$--$70$ to $50$--$85$~K), above both the field ($0.934$--$0.986$) and the conformal bath. The local slope rises from $0.626$ at $47$~K to $0.637$~$\mu\Omega$\,cm/K at $77$~K, where the field predicts it to rise on cooling by $0.6$--$3.6\%$; a superlinear term is therefore present below $85$~K, and it grows above (slope $0.83$ at $230$~K), as is known for overdoped LSCO~\cite{Cooper2009}. Windows reaching down to $30$--$35$~K give $n=1.00$--$1.01$, lowered by the steeper slope near $T_{c}$ ($0.642$ at $35$~K, $1.11$ at $24$~K), which we attribute to superconducting fluctuations. A fit of the drift over $40$--$263$~K with a $T^{2}$ or $T^{3}$ background returns anything from $-0.19$ to $+0.50$ per decade, so it is set by the background model. The $16$~T curve has a lower slope at low $T$ ($0.609$ at $20$--$40$~K), consistent with magnetoresistance~\cite{GiraldoGallo2018}.

Field-induced magnetoresistance, which in overdoped cuprates scales as $\sqrt{(\alpha T)^{2}+(\gamma_{H}H)^{2}}$~\cite{GiraldoGallo2018} and lowers the apparent slope at low $T$, superconducting fluctuations near $T_{c}(H)$, and the superlinear term enter at the same $1$--$5\%$ level as the predicted curvature. The published curves therefore neither establish nor exclude the drift; taken at face value, the LSCO curves lean against it and the Nd-LSCO curve toward it. The decisive measurement is a zero-field-extrapolated $\rho(T)$ over $1$--$100$~K on one crystal at $p^{*}$, with the magnetoresistance removed point by point.

\section{Other baths and limitations}
\label{sec:Sother}
\begin{figure}[htb]
\includegraphics[width=\columnwidth]{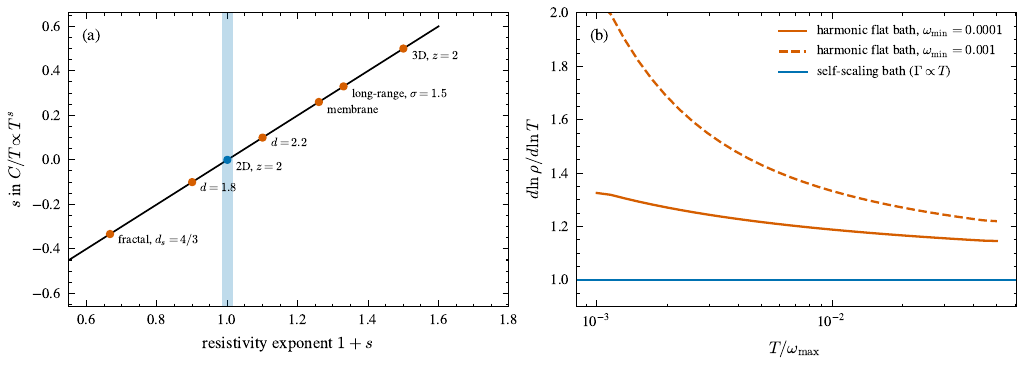}
\caption{(a)~A single local bath with spectrum $\propto\Omega^{s}$ gives $C/T\propto T^{s}$ and $\rho\propto T^{1+s}$. Points: fractal network ($d_{s}=4/3$), fractional dimension $d=1.8$ and $2.2$ (overdamped, $z=2$), anharmonic membrane, fractional Laplacian ($\sigma=1.5$), three-dimensional $z=2$; only $d=2$, $z=2$ meets the measured $T$-linear resistivity. (b)~Logarithmic slope of the rate for a harmonic bath with a fixed flat spectrum above $\omega_{\min}$ (in units of the cutoff): it stays above $1$. A bath whose infrared scale is $\propto T$ gives exactly $1$.}
\label{fig:exponents}
\end{figure}

A bath with a power-law spectrum fixes the exponents of both laws. Let $\aF\propto\Omega^{s}$ above $T$. The rate at $\varepsilon=0$ is $\int d\Omega\,\Omega^{s}/\sinh(\Omega/T)\propto T^{1+s}$, and the fixed-bath free energy of Sec.~\ref{sec:Sbath} gives $C/T\propto T^{s}$ (a logarithm at $s=0$). For an overdamped field $D^{-1}=q^{\sigma}+c|\Omega|$ in $d$ dimensions, $\int d^{d}q\,\Im D\propto\Omega^{(d-\sigma)/\sigma}$; for a propagating mode with $\omega\propto q^{z}$ the density of states gives $s=d/z-1$; on a fractal, $d$ is replaced by the spectral dimension $d_{s}$~\cite{AlexanderOrbach1982}. Flexural modes of a free membrane have $z=2$ and $s=0$ but are harmonic; thermal anharmonicity makes them anomalous with $s\approx0.26$~\cite{NelsonPeliti1987}, and electron damping turns them into an overdamped $z=4$ mode with $s=-\frac12$. A relaxational central-peak mode~\cite{HalperinVarma1976} with a fixed defect rate is a Fermi liquid below that rate. Free two-level systems with a flat distribution of splittings give exactly the conformal shape, $\aF\propto\tanh(\Omega/2T)$, so they belong to the conformal point; their own entropy is $C\propto T$, not a logarithm. A harmonic bath with a fixed infrared scale gives $\rho\propto T\ln T$, with a logarithmic slope of $1.15$--$1.3$ at $\omega_{\min}=10^{-4}$ and $1.23$--$1.9$ at $10^{-3}$ (Fig.~\ref{fig:exponents}b); a $T$-linear rate needs the bath's relaxation rate to scale with $T$. With $z=2$, a logarithmic $C/T$ together with a $T$-linear resistivity needs two-dimensional critical fluctuations. CeCu$_{6-x}$Au$_{x}$ has two-dimensional critical fluctuations~\cite{Stockert1998}, while three-dimensional spin-density-wave cases show $C/T=\gamma_{0}-a\sqrt{T}$ and $\rho\propto T^{3/2}$~\cite{Lohneysen2007}.

A strange-metal bath must be critical at $p^{*}$, have a flat ($s=0$) spectrum whose relaxation rate scales with $T$, and couple locally. Phonons and polarons are not critical at $p^{*}$ and are not marginal. Charge-order fluctuations at finite wave vector produce hot spots, and their measured relaxation rate, $0.65$--$0.9\,\kB T$~\cite{Arpaia2023}, is faster than LSCO's $r\approx0.3$--$0.55$ unless the damping is larger than in the simplest reading. Spin fluctuations relax at $2.9\,\kB T$~\cite{Radaelli2026}, which would give $\beta\approx1$ and $R_{\rm opt}\approx2.3$, against $8.0$--$9.1$. Pair fluctuations cannot be the main bath, because the $T$-linear resistivity persists from about $1$ to $300$~K and heavy fermions such as YbRh$_{2}$Si$_{2}$ show the same phenomenology far above $T_{c}$; a subdominant pair share, up to about $30\%$, is not excluded. Charge-$4e$ modes couple to four fermions and, by the power counting of Ref.~\cite{SinWang2025}, give no $T$-linear rate. Fractal, fractional-dimensional, long-range elastic and membrane modes fail the exponent test above. Octahedral-tilt modes of LSCO are zone-boundary modes with quadratic coupling, absent in Bi- and Tl-based cuprates with the same strange metal. A uniform coupling to a $q=0$ boson as the main channel is not $T$-linear in two dimensions~\cite{DellAnnaMetzner2007,MaslovYudsonChubukov2011} and gives $\eta\ll1$.

The analysis assumes that the bath is local, as motivated by ADMR~\cite{Grissonnanche2021}. Identifying the fitted photoemission exponent with Eq.~(\ref{eq:beta}) assumes a fitted power near $\frac12$, and the photoemission $\beta$ is measured in Bi-based cuprates, not at $p^{*}$ in La-based ones. The two determinations of $\lambda$ in LSCO (optical slope $0.40$, optical mass logarithm $0.46$) differ by $15\%$, although in a local marginal bath they should coincide, and both routes to $r$ use Michon et al.'s zero-frequency value. The entropy share $\eta$ carries the band systematic of Sec.~\ref{sec:Sdata}. The floor on $\beta$ is a large-$M$ result, and $1/M$ corrections were not computed. The reservoir route is a leading-logarithm argument with an Ohmic reservoir that carries no logarithm of its own. We do not identify the microscopic field.


\end{document}